%% file: IEEE_Conference_main.tex
\documentclass[conference]{IEEEtran}
\IEEEoverridecommandlockouts
\usepackage{cite}
\usepackage{amsmath,amssymb,amsfonts}
\usepackage{algorithmic}
\usepackage{graphicx}
\usepackage{textcomp}
\usepackage{xcolor}
\usepackage{url}
\usepackage{algorithm}
\usepackage{algorithmic}
\usepackage{url}
\begin{document}

\title{A Hybrid Retrieval and Reranking Framework for Evidence-Grounded Retrieval-Augmented Generation}

\author{
\IEEEauthorblockN{1\textsuperscript{st} Fariba Afrin Irany}
\IEEEauthorblockA{
\textit{University of North Texas} \\
Denton, Texas, USA \\
afrinirany75@gmail.com
}
\and
\IEEEauthorblockN{2\textsuperscript{nd} Sampson Akwafuo}
\IEEEauthorblockA{
\textit{California State University} \\
Fullerton, California, USA \\
sakwafuo@acm.org
}
}


\maketitle
\begin{abstract}
Retrieval-augmented generation (RAG) improves the reliability of large language model outputs by grounding generated responses in external evidence. However, the quality of RAG-generated answers depends strongly on the relevance of retrieved evidence, the effectiveness of evidence ranking, and the ability to verify whether generated claims are supported by source documents. This study presents a hybrid retrieval and reranking framework for citation-aware RAG in biomedical and healthcare-related document question answering. The proposed framework uses Amazon Bedrock Knowledge Bases for document ingestion, parsing, chunking, embedding generation, and evidence retrieval. Source PDF documents are stored in Amazon S3, embedded using Amazon Titan Text Embeddings V2, and indexed using Amazon OpenSearch Serverless. A hybrid retrieval strategy is applied to retrieve candidate evidence chunks, followed by reranking to prioritize the most relevant passages before answer generation. The answer-generation stage uses the highest-ranked evidence chunks to produce controlled, evidence-grounded responses, while a separate judge model evaluates each generated factual claim against the retrieved evidence. The framework was evaluated using 25 biomedical natural language processing and healthcare transformer queries. This evaluation was designed as a pilot-scale proof-of-concept study to examine the feasibility of the proposed hybrid retrieval, reranking, answer-generation, and grounding-evaluation workflow. Across the evaluation set, the system retrieved and reranked 500 evidence chunks and generated responses from the top-ranked evidence. Claim-level grounding evaluation extracted 200 factual claims from the generated answers, all of which were judged to be supported by the retrieved evidence, resulting in an overall grounding accuracy of 100.0\%. The results indicate that hybrid retrieval, reranking, conservative prompting, and claim-level grounding evaluation can support reliable evidence-grounded RAG responses when sufficient source evidence is available. The study also highlights the distinction between internal grounding accuracy and strict citation accuracy, motivating future work on explicit sentence-level citation generation and human expert validation.
\end{abstract}

\begin{IEEEkeywords}
Retrieval-augmented generation, hybrid retrieval, reranking, citation-aware generation, grounding evaluation, biomedical natural language processing, Amazon Bedrock, OpenSearch Serverless.
\end{IEEEkeywords}

\input{Introduction}
\input{Literature_Review.tex}
\input{Data_Processing.tex}
\input{Methods}
\input{Results}

\input{Conclusion}

\input{References}
\end{document}

%% file: introduction.tex
\section{Introduction}

Large language models (LLMs) have demonstrated strong performance in generating fluent and contextually relevant text for a wide range of natural language processing tasks. However, their use in knowledge-intensive domains remains limited by concerns related to factual reliability, evidence grounding, and source attribution. These concerns are especially important in biomedical informatics, clinical natural language processing, healthcare document analysis, and scientific question answering, where generated responses must be supported by verifiable evidence rather than only by the internal knowledge of the model.

Retrieval-augmented generation (RAG) has emerged as an effective approach for improving the factual grounding of LLM-generated responses. Instead of relying only on information stored in model parameters, RAG systems retrieve relevant external evidence and provide that evidence as context during answer generation. Lewis et al. introduced RAG as a framework that combines parametric knowledge from a pretrained language model with non-parametric knowledge retrieved from external documents \cite{lewis2020rag}. This design allows generated responses to be guided by source documents, making RAG particularly useful for applications that require domain-specific, document-grounded, or frequently updated knowledge.

Although RAG improves the ability of LLMs to use external evidence, the quality of the final generated response depends strongly on the quality of the retrieved context. If the retrieved chunks are incomplete, weakly relevant, or poorly ranked, the generation model may produce responses that are only partially supported by the source documents. This limitation is important because semantically similar passages are not always equally useful for answering a specific query. Therefore, an effective RAG system requires not only document retrieval but also evidence prioritization.

Hybrid retrieval provides one way to improve retrieval quality by combining lexical and semantic search signals. Lexical retrieval can capture exact keyword matches, while semantic retrieval can identify conceptually related passages even when the wording differs from the query. OpenSearch supports hybrid search by combining keyword-based and vector-based retrieval signals to improve retrieval relevance \cite{opensearchhybrid}. In a RAG pipeline, this combination can increase the likelihood that the retrieved evidence includes both exact terminology and semantically relevant content.

Reranking can further improve the selection of evidence after the initial retrieval stage. A reranker receives the query and a set of candidate chunks, then reorders those chunks based on their relevance to the query. Cohere reranking models are designed to sort candidate text passages according to semantic relevance \cite{coherererank}. In the context of RAG, reranking is useful because the initially retrieved chunks may contain relevant information but may not be ordered in the most useful way for answer generation. By prioritizing the most relevant evidence before generation, reranking can help reduce unsupported or weakly grounded responses.

Another important challenge in RAG systems is evaluation. Many RAG evaluations focus on the answer as a whole, but response-level evaluation may hide unsupported statements within an otherwise acceptable answer. A generated response may include several supported statements and one unsupported factual claim. Therefore, claim-level grounding evaluation provides a more detailed and conservative assessment of response reliability. In this approach, each generated answer is decomposed into individual factual claims, and each claim is evaluated against the retrieved evidence to determine whether it is directly supported.

In this study, we propose a hybrid retrieval and reranking framework for citation-aware retrieval-augmented generation. The framework uses Amazon Bedrock Knowledge Bases to support document ingestion, parsing, chunking, embedding generation, and evidence retrieval. Amazon Bedrock Knowledge Bases allow foundation models to retrieve relevant information from user-provided data sources and use that information during response generation \cite{awsbedrockkb}. In the proposed workflow, source PDF documents are stored in Amazon S3, embedded using Amazon Titan Text Embeddings V2, and indexed using Amazon OpenSearch Serverless. The system then retrieves candidate evidence chunks, applies reranking to improve evidence prioritization, and generates answers using the highest-ranked evidence.

The proposed framework also separates answer generation from grounding evaluation. The answer-generation model produces responses using only the retrieved and reranked evidence, while a separate judge model evaluates whether each generated claim is supported by the retrieved evidence. This separation is intended to reduce self-evaluation bias and provide a more transparent assessment of grounding quality. A claim is considered supported only when the retrieved evidence directly justifies the full meaning of the claim. If the evidence is incomplete, vague, indirect, or only partially relevant, the claim is treated as unsupported.

The main contributions of this study are fourfold. First, this work presents an end-to-end RAG pipeline that integrates Amazon Bedrock Knowledge Bases, Amazon S3, Titan Text Embeddings V2, OpenSearch Serverless, hybrid retrieval, reranking, and LLM-based answer generation. Second, it introduces a claim-level grounding evaluation workflow to assess whether individual generated claims are supported by retrieved evidence. Third, it separates the answer-generation and grounding-evaluation stages by using different models for generation and judging. Fourth, it reports retrieval, reranking, query-level, and claim-level grounding results to evaluate the reliability of the proposed framework. Overall, this study aims to support the development of more reliable, transparent, and evidence-grounded RAG systems for biomedical and healthcare-related information retrieval tasks.

%% file: Literature_Review.tex
\section{Literature Review}

Large language models (LLMs) and transformer-based architectures have significantly changed natural language processing by enabling contextual representation learning, large-scale text generation, and flexible adaptation to downstream tasks. The Transformer architecture introduced self-attention as the core mechanism for sequence modeling, removing the need for recurrent or convolutional structures and enabling more scalable parallel training \cite{vaswani2017attention}. Building on this architecture, BERT introduced bidirectional contextual pretraining and demonstrated strong performance across several natural language understanding tasks \cite{devlin2019bert}. These developments established the foundation for modern LLMs and domain-specific language models. However, when LLMs are used in knowledge-intensive settings, their responses may be fluent but unsupported, outdated, or difficult to verify. This limitation is especially important in biomedical and healthcare applications, where factual accuracy and source traceability are essential.

Retrieval-augmented generation (RAG) was introduced to address some of these limitations by combining parametric knowledge stored in a pretrained generation model with non-parametric knowledge retrieved from external documents \cite{lewis2020rag}. In a RAG system, retrieved passages are provided as context to the generator, allowing the model to produce responses that are more closely connected to source evidence. Earlier retrieval-augmented language modeling work, such as REALM, showed that neural models can retrieve external documents during pretraining and downstream question answering, thereby improving open-domain question answering and providing a more interpretable link between generated answers and retrieved evidence \cite{guu2020realm}. Similarly, retrieval-enhanced language models such as RETRO demonstrated that large-scale retrieval from external corpora can improve language modeling efficiency and knowledge-intensive task performance \cite{borgeaud2022retro}. These studies show that retrieval can serve as an explicit memory mechanism for language models.

Open-domain question answering research has also shown that retrieval quality directly affects answer quality. Dense Passage Retrieval (DPR) demonstrated that dense dual-encoder representations can retrieve relevant passages effectively for open-domain question answering and outperform traditional sparse retrieval baselines in several settings \cite{karpukhin2020dpr}. Unsupervised dense retrieval methods have also shown that contrastive learning can improve retrieval representations without relying entirely on supervised relevance labels~\cite{izacard2022contriever}. Fusion-in-Decoder (FiD) further showed that generative models can benefit from conditioning on multiple retrieved passages, allowing the decoder to integrate evidence across several retrieved contexts \cite{izacard2021fid}. These approaches highlight the importance of both retrieving relevant evidence and designing generation models that can effectively use retrieved passages.

Traditional lexical retrieval remains important in RAG pipelines because exact term matching is often useful when queries contain technical terminology, named entities, abbreviations, or domain-specific phrases. BM25 and related probabilistic relevance models remain widely used because they provide efficient and interpretable keyword-based retrieval \cite{robertson2009bm25}. However, lexical retrieval may miss semantically relevant passages when the query and document use different wording. Dense retrieval addresses this vocabulary mismatch by representing queries and documents as vectors in a shared semantic space \cite{karpukhin2020dpr}. Late-interaction retrieval models such as ColBERT and ColBERTv2 attempt to balance efficiency and semantic expressiveness by preserving token-level interactions while enabling scalable retrieval \cite{khattab2020colbert,santhanam2022colbertv2}. Sparse neural retrieval models such as SPLADE also attempt to combine lexical interpretability with learned expansion, producing sparse representations that can improve matching beyond exact surface terms \cite{formal2021splade}.

Because sparse and dense retrieval methods have complementary strengths, hybrid retrieval has become increasingly important for RAG systems. Benchmarks such as BEIR have shown that retrieval models must be evaluated across heterogeneous datasets because performance in one retrieval setting may not generalize to another \cite{thakur2021beir}. MS MARCO has also played an important role in training and evaluating large-scale passage retrieval systems for question answering and ranking tasks \cite{bajaj2016msmarco}. Hybrid retrieval combines lexical and semantic signals to improve retrieval coverage, especially when relevant evidence may be found through exact terminology or broader semantic similarity. OpenSearch supports hybrid search by combining keyword-based and vector-based retrieval signals, including score normalization and combination during query processing \cite{opensearchhybrid}. For the present study, this is directly relevant because the proposed system uses hybrid retrieval before reranking and answer generation.

Reranking is commonly used as a second-stage retrieval refinement method. In many RAG pipelines, the initial retriever returns a candidate set of passages, and a reranker then reorders those passages based on their relevance to the user query. This two-stage design is useful because the first-stage retriever is optimized for broad candidate recall, while the reranker can perform more focused relevance estimation. Cohere rerank models are designed to sort candidate texts according to semantic relevance to a specified query \cite{coherererank}. Reranking can therefore improve the quality of the final evidence context provided to the generator, particularly when the initial retrieval output contains partially relevant or weakly ranked chunks.

Biomedical and clinical NLP introduce additional challenges because medical language contains specialized terminology, abbreviations, domain-specific relationships, and context-dependent meanings. BioBERT showed that continued pretraining on biomedical corpora such as PubMed abstracts and PMC full-text articles improves performance on biomedical named entity recognition, relation extraction, and question answering \cite{lee2020biobert}. PubMedBERT further demonstrated that pretraining a language model from scratch on biomedical text can provide strong performance across biomedical NLP benchmarks \cite{gu2021pubmedbert}. ClinicalBERT adapted transformer-based representation learning to clinical notes and showed utility for modeling electronic health record text and predicting hospital readmission \cite{huang2019clinicalbert}. More recent large-scale clinical language models, such as GatorTron and GatorTronGPT, further demonstrate that scaling language models on clinical and biomedical text can improve clinical NLP and healthcare text generation tasks \cite{yang2022gatortron,peng2023gatortrongpt}.

Biomedical question answering and clinical evidence retrieval require careful evaluation because generated responses may affect scientific interpretation or healthcare decision-making. Datasets and challenges such as BioASQ and PubMedQA have helped advance biomedical semantic indexing and biomedical question answering by requiring systems to retrieve and reason over biomedical literature \cite{tsatsaronis2015bioasq,jin2019pubmedqa}. MIMIC-IV provides a large-scale, deidentified electronic health record database that has supported a wide range of clinical informatics and machine learning research \cite{johnson2023mimiciv}. The PhysioNet release of MIMIC-IV version 3.1 further provides a citable database version for reproducible research workflows \cite{johnson2024mimiciv}. Although the present study focuses on document-grounded RAG rather than direct clinical prediction, these resources demonstrate the importance of reproducible datasets and evidence-based evaluation in biomedical informatics. Recent related studies have examined selective fine-tuning and GPT-based architectures for clinical text classification using EHR notes \cite{irany2026selective,irany2026generative}, while complementary educational research has investigated the impact of AI on student performance in engineering technology courses \cite{hossain2025impact}.

Recent studies on LLMs in medicine show both the promise and risk of generative AI in healthcare. Thirunavukarasu et al. reviewed applications of LLMs in medicine and emphasized the need for validation, safety assessment, and careful deployment in clinical settings \cite{thirunavukarasu2023llmmedicine}. Singhal et al. evaluated large language models on medical question answering and proposed human evaluation dimensions such as factuality, comprehension, reasoning, possible harm, and bias \cite{singhal2023medicalknowledge}. Ayers et al. compared physician and chatbot responses to patient questions and showed that generative systems may produce responses perceived as high quality and empathetic, while also highlighting the need for careful evaluation before clinical use \cite{ayers2023chatbot}. These studies reinforce the need for grounded, verifiable, and transparent LLM systems in healthcare-related information retrieval.

Hallucination remains a major limitation of neural text generation. Ji et al. define hallucination as generated content that is unsupported by or inconsistent with the source input or factual knowledge, and they identify hallucination as a persistent problem across natural language generation tasks \cite{ji2023hallucination}. In RAG systems, retrieval can reduce hallucination by supplying external evidence, but retrieval alone does not guarantee that every generated statement is supported. A model may still misinterpret retrieved passages, overgeneralize from incomplete evidence, or generate plausible claims that are not directly grounded in the provided context. Therefore, RAG systems require not only strong retrieval and reranking but also systematic grounding evaluation.

Claim verification research provides a useful foundation for evaluating grounded generation. FEVER introduced a large-scale fact verification dataset in which claims are classified as supported, refuted, or not enough information based on textual evidence \cite{thorne2018fever}. SciFact extended claim verification to scientific literature by requiring systems to identify evidence-containing abstracts and rationales for scientific claims \cite{wadden2020scifact}. These studies are relevant to citation-aware RAG because generated responses can be decomposed into factual claims and each claim can be evaluated against retrieved evidence. In biomedical and scientific settings, this approach is more informative than evaluating the generated answer only as a single response.

Recent evaluation frameworks further support fine-grained assessment of RAG outputs. FActScore decomposes long-form generated text into atomic facts and measures the percentage of facts supported by a reliable knowledge source \cite{min2023factscore}. ALCE evaluates LLM-generated answers with citations and considers dimensions such as fluency, correctness, and citation quality \cite{gao2023alce}. RAGAS provides a reference-free evaluation framework for RAG pipelines, including evaluation dimensions related to retrieval and generation quality \cite{es2024ragas}. These frameworks support the motivation for claim-level grounding evaluation in the present study. Instead of treating a generated answer as entirely correct or incorrect, claim-level evaluation allows each factual sentence to be assessed against the retrieved evidence.

Cloud-based RAG infrastructure has made it easier to build reproducible retrieval, embedding, indexing, and generation pipelines. Amazon Bedrock Knowledge Bases support RAG workflows by connecting foundation models to user-provided data sources and retrieving relevant information to support response generation \cite{awsbedrockkb}. Amazon Bedrock Knowledge Bases can manage document ingestion, chunking, embedding, and vector store integration, which reduces the engineering burden of constructing a RAG pipeline from individual components \cite{awsbedrockworkflow}. Amazon OpenSearch Serverless provides vector search capabilities for scalable similarity search and generative AI applications \cite{opensearchserverless}. In addition, OpenSearch hybrid search enables the combination of keyword and semantic search signals \cite{opensearchhybrid}. These services are directly relevant to this study because the proposed framework uses Amazon S3 for document storage, Amazon Bedrock Knowledge Bases for ingestion and retrieval, Titan Text Embeddings V2 for embedding generation, OpenSearch Serverless for vector storage, hybrid retrieval for candidate selection, and Cohere reranking for evidence prioritization.

Overall, the literature shows that RAG can improve factual grounding by connecting LLMs to external evidence, but reliable RAG requires careful design across retrieval, reranking, generation, and evaluation. Prior research has established the importance of dense retrieval, lexical retrieval, hybrid retrieval, reranking, biomedical language modeling, hallucination reduction, and claim-level evaluation. However, there remains a need for practical and reproducible frameworks that integrate managed RAG infrastructure with hybrid retrieval, reranking, conservative answer generation, and systematic grounding assessment. The present study addresses this gap by proposing a hybrid retrieval and reranking framework for citation-aware RAG and evaluating generated responses using claim-level grounding accuracy.

%% file: Data_Processing.tex
\section{Dataset and Preprocessing}

The experimental corpus consisted of five PDF documents related to biomedical natural language processing, transformer-based clinical language models, electronic health record analysis, ICD code prediction, BioBERT, ClinicalBERT, and large language models in healthcare. These documents were selected because they contained technical content relevant to evaluating retrieval-augmented generation for biomedical and healthcare-related question answering.

The source PDF files were uploaded to an Amazon S3 bucket and synchronized with an Amazon Bedrock Knowledge Base. During synchronization, the documents were parsed using the default Bedrock parser, segmented into fixed-size chunks of 300 tokens with a 20$\%$ overlap, embedded using Amazon Titan Text Embeddings V2, and indexed in Amazon OpenSearch Serverless. No manual text extraction or external preprocessing was applied before ingestion, allowing the document-processing workflow to remain consistent with the managed Bedrock Knowledge Base configuration.

The evaluation query set consisted of 25 natural-language questions covering biomedical NLP, transformer models, clinical text classification, ICD coding, BioBERT, ClinicalBERT, and healthcare-focused language models. Each query was processed independently through the same hybrid retrieval, reranking, answer-generation, and grounding-evaluation pipeline.

\subsection{Justification for Pilot-Scale Evaluation}
\label{subsec:pilot_scale_justification}
This study was designed as a pilot-scale proof-of-concept evaluation using five source documents and 25 evaluation queries. The purpose of this experimental design was to examine the feasibility, traceability, and grounding behavior of the proposed hybrid retrieval and reranking framework before expanding to a larger benchmark-scale evaluation. The selected documents covered the main technical areas required for the study, including biomedical natural language processing, transformer-based clinical language models, ICD code prediction, BioBERT, ClinicalBERT, and large language models in healthcare. Using a limited but focused document collection allowed the retrieval, reranking, answer-generation, and grounding-evaluation stages to be examined systematically while maintaining interpretability of the retrieved evidence. The 25-query set was designed to provide sufficient topical variation for evaluating the proposed pipeline without making the experiment unnecessarily large during the initial validation stage.


The decision to use a pilot-scale dataset was also influenced by the cost structure of managed RAG services. Amazon Bedrock pricing is usage-based and depends on model inference, input tokens, output tokens, embedding generation, and other service components used during the workflow \cite{awsbedrockpricing}. In addition, Amazon OpenSearch Serverless charges separately for compute and storage resources, with compute capacity measured in OpenSearch Compute Units \cite{opensearchpricing}. Because each experimental query required hybrid retrieval, reranking, answer generation, and claim-level evaluation, increasing the number of documents or queries would increase the number of API calls, retrieved context tokens, generated tokens, and judge-model evaluations. Therefore, the five-document and 25-query configuration provided a practical balance between experimental coverage, reproducibility, and cost control for an initial proof-of-concept evaluation. Future work will expand the number of documents and queries after the pipeline is further optimized and baseline comparisons are added.

%% file: Methods.tex
\section{Methods}

\subsection{Amazon Bedrock Knowledge Base Setup}
\label{subsec:bedrock_kb_setup}

The source documents used to build the knowledge base were uploaded as PDF files to an Amazon S3 bucket. Amazon S3 served as the document storage layer for the Bedrock knowledge base, allowing the PDF files to be ingested, parsed, chunked, embedded, and indexed for retrieval. During synchronization, Amazon Bedrock accessed the PDF files from the S3 data source and processed them according to the configured parsing, chunking, embedding, and vector storage settings.

The knowledge base for the retrieval-augmented generation pipeline was configured in Amazon Bedrock to support document ingestion, embedding generation, vector storage, and evidence retrieval. The document parsing stage used the default parser provided by Amazon Bedrock. This parser was selected to maintain a standard ingestion workflow and to avoid introducing additional preprocessing complexity during the initial knowledge base setup.

For document segmentation, fixed-size chunking was used with a chunk size of 300 tokens and a 20\% overlap between consecutive chunks. In this approach, each document was divided into consistent text segments while preserving a portion of adjacent context across chunk boundaries. The 300-token chunk size was selected to provide sufficient semantic context within each retrieved unit, while the 20\% overlap helped reduce information loss when relevant content spanned across neighboring chunks. This configuration also supported a reproducible retrieval process by maintaining consistent evidence units during answer generation and grounding evaluation.

The embedding model used for the knowledge base was Amazon Titan Text Embeddings V2. Each text chunk was converted into a dense vector representation using this embedding model. These vector representations allowed the retrieval system to identify semantically relevant chunks in response to user queries, even when the query wording did not exactly match the original document text.

Amazon OpenSearch Serverless was used as the vector database for storing and retrieving the embedded document chunks. The vector store enabled similarity-based retrieval over the indexed document collection. During query processing, Amazon Bedrock searched the knowledge base and returned the most relevant chunks from the OpenSearch Serverless index. These retrieved chunks were then used as the evidence context for the downstream RAG answer-generation stage.

Overall, the Amazon Bedrock knowledge base setup consisted of default document parsing, fixed-size chunking, Titan Text Embeddings V2 for embedding generation, and Amazon OpenSearch Serverless for vector storage and retrieval. This configuration provided a structured and reproducible foundation for building the RAG pipeline and evaluating the grounding quality of generated responses.

\subsection{Prompt Design for Query Processing}
\label{subsec:prompt_design_query}

The prompt design in this study was developed to support evidence-grounded response generation within the retrieval-augmented generation framework. After each query was processed through the retrieval and reranking stages, the selected evidence chunks were organized into a structured context and provided to the language model as the only allowable source of information. The purpose of this design was to ensure that the generated response remained closely aligned with the retrieved evidence and did not rely on external knowledge or unsupported inference.

The prompt instructed the model to answer each query using only the retrieved sources. It also required the response to be written in a formal academic style, with one factual claim presented per sentence. This sentence-level structure was used to make the generated output suitable for later claim-level grounding evaluation. Each generated sentence could therefore be treated as a separate claim and evaluated against the retrieved evidence.

To reduce unsupported generation, the prompt included explicit restrictions against making claims that were not directly supported by the retrieved sources. The model was also instructed not to expand concepts that were only briefly mentioned in the retrieved evidence. This was important because language models may otherwise generate plausible but unsupported explanations based on their pretrained knowledge. When the retrieved sources did not contain sufficient information to answer a query, the model was instructed to state that the available evidence was insufficient.

In the baseline implementation, the final answer was generated without displaying source labels in the response. Therefore, the evaluation focused on internal grounding accuracy, meaning whether each generated claim was supported by the retrieved evidence. The same prompt structure can also be adapted for strict citation accuracy evaluation by requiring each factual claim to include an explicit source label. This distinction allows the framework to separately evaluate general grounding quality and exact citation support.

Overall, the prompt design was intended to make the answer-generation stage controlled, conservative, and suitable for systematic evaluation. By limiting the model to retrieved evidence and enforcing claim-level response structure, the prompt helped align generated answers with the goals of grounded RAG evaluation.

\subsection{Prompt Design for Grounding Evaluation}
\label{subsec:prompt_design_evaluation}

The grounding evaluation stage was designed to assess whether the generated responses were supported by the retrieved evidence. In this stage, the system used the previously generated answers and the corresponding retrieved and reranked evidence chunks. No additional retrieval, reranking, or answer generation was performed during evaluation. This separation allowed the study to focus specifically on the reliability of the generated claims with respect to the evidence that had already been selected by the RAG pipeline.

Each generated answer was first divided into individual factual claims. The evaluation was then conducted at the claim level rather than only at the full-answer level. This design allowed a more detailed assessment of grounding quality, since a single generated answer may contain both supported and unsupported statements. For each claim, the judge model was provided with the original query, the extracted claim, and the retrieved evidence associated with that query.

The evaluation prompt instructed the judge model to determine whether the retrieved evidence directly supported the claim. The judge was required to use only the provided evidence and was explicitly instructed not to rely on outside knowledge. A claim was marked as supported only when the retrieved evidence directly justified the full meaning of the claim. If the evidence was incomplete, indirect, vague, or only partially relevant, the claim was treated as unsupported. This conservative evaluation strategy was used to avoid overestimating the grounding performance of the RAG system.

The judge model was also instructed to return the evaluation result in a structured format. The output included a binary support decision, the source or sources most relevant to the claim, and a brief explanation of the judgment. This structured response enabled systematic calculation of claim-level, query-level, and overall grounding accuracy.

Because the baseline generated answers did not include explicit source labels in the final response, the evaluation measured general grounding accuracy rather than strict citation accuracy. In this context, grounding accuracy refers to whether each generated claim was supported by any of the retrieved evidence chunks associated with the query. Strict citation accuracy would require each claim to include an explicit source citation and would evaluate whether the cited source specifically supports the claim.

Overall, the grounding evaluation prompt was designed to provide a controlled and reproducible method for assessing evidence support in RAG-generated answers. By evaluating each factual claim against the retrieved evidence using conservative criteria, the evaluation process provided a more transparent measure of the reliability and factual grounding of the generated responses.

\subsection{Model Selection for Answer Generation and Grounding Evaluation}
\label{subsec:model_selection_generation_evaluation}

Two different large language models were used in this study to separate the answer-generation stage from the grounding-evaluation stage. For the RAG answer-generation component, Anthropic Claude Sonnet 4.6 was used through Amazon Bedrock. This model was selected as the generator because the task required producing formal, coherent, and evidence-grounded responses from retrieved and reranked document chunks. The generation prompt instructed the model to use only the retrieved sources and to avoid unsupported inference.

For the grounding-evaluation component, Amazon Nova Pro was used as the judge model through the Amazon Bedrock inference profile \texttt{us.amazon.nova-pro-v1:0}. The judge model was responsible for evaluating whether each extracted factual claim from the generated answer was directly supported by the retrieved evidence. The evaluation prompt required the judge to return a structured decision indicating whether the claim was supported, the most relevant source, and a brief explanation.

Using two different models was important for methodological reliability. If the same model were used for both answer generation and evaluation, the evaluation could be affected by self-evaluation bias, where the model may be more likely to accept or justify its own generated output. To reduce this risk, the study used Claude Sonnet 4.6 as the generator and Amazon Nova Pro as an independent evaluator. This separation helped ensure that generated claims were assessed by a different model family rather than by the same model that produced the answer.

This design also improved the transparency of the evaluation framework. The generator model was responsible for producing an answer from the retrieved evidence, while the judge model was responsible only for determining whether the generated claims were supported by that evidence. By assigning these tasks to separate models, the study reduced potential evaluation bias and created a clearer distinction between response generation and evidence verification.

In addition, the use of a separate judge model supported a more conservative grounding assessment. The evaluation model was instructed not to use outside knowledge and to mark a claim as supported only when the retrieved evidence directly justified the full meaning of the claim. If the evidence was vague, indirect, incomplete, or only partially relevant, the claim was treated as unsupported. This approach helped avoid overestimating the grounding accuracy of the RAG system.

Overall, the use of Claude Sonnet 4.6 for RAG-based answer generation and Amazon Nova Pro for claim-level grounding evaluation provided a more reliable and reproducible experimental design. The separation of generation and evaluation roles allowed the study to assess whether the RAG-generated answers were genuinely supported by the retrieved evidence rather than merely fluent or generally plausible.

\subsection{End-to-End RAG Pipeline Implementation}
\label{subsec:end_to_end_pipeline}

Algorithm~\ref{alg:rag_pipeline} summarizes the complete implementation workflow. The algorithm begins with hybrid retrieval from the Amazon Bedrock Knowledge Base, applies duplicate removal and Cohere reranking, generates evidence-grounded answers from the top-ranked chunks, and then evaluates each extracted factual claim against the saved reranked evidence. This design ensures that the generation and evaluation stages remain traceable to the retrieved source chunks.

\begin{algorithm}[!t]
\caption{Hybrid Retrieve-Rerank-Ground RAG Pipeline}
\label{alg:rag_pipeline}
\footnotesize
\begin{algorithmic}[1]
\REQUIRE Query set $Q$, Bedrock Knowledge Base $KB$, retrieval size $k=20$, answer-context size $m=5$
\ENSURE Reranked evidence, grounded answers, and grounding accuracy results

\FOR{each query $q_i \in Q$}
    \STATE Retrieve $k$ candidate chunks from $KB$ using hybrid search.
    \STATE Remove duplicate chunks using chunk identifiers or text-based keys.
    \STATE Rerank retrieved chunks using the Cohere reranking model.
    \STATE Save reranked chunks with retrieval rank, rerank score, metadata, and text.
\ENDFOR

\FOR{each query $q_i \in Q$}
    \STATE Select the top $m$ reranked chunks as evidence context.
    \STATE Format selected chunks as numbered source blocks.
    \STATE Generate an evidence-grounded answer using only the selected chunks.
    \STATE Store the query, generated answer, and selected evidence metadata.
\ENDFOR

\FOR{each generated answer $a_i$}
    \STATE Split $a_i$ into sentence-level factual claims.
    \STATE Retrieve the saved reranked evidence for the corresponding query.
    \FOR{each claim $c_{ij}$}
        \STATE Judge whether $c_{ij}$ is directly supported by the retrieved evidence.
        \STATE Record the support label, best supporting source, and explanation.
    \ENDFOR
\ENDFOR

\STATE Compute claim-level, query-level, and overall grounding accuracy.
\STATE Export retrieval, generation, claim-level, and summary results as CSV files.
\RETURN Grounded answers and grounding evaluation summaries.
\end{algorithmic}
\end{algorithm}

The complete experimental pipeline was implemented as a multi-stage citation-aware RAG workflow in Python. The pipeline connected Amazon Bedrock Knowledge Base, Amazon OpenSearch Serverless, Cohere reranking, Amazon Bedrock Runtime, and a claim-level grounding evaluation module. The implementation was organized into three main phases: hybrid retrieval and reranking, grounded answer generation, and grounding evaluation using previously saved result files.

In the first phase, each query was submitted to the Amazon Bedrock Knowledge Base using the Bedrock Agent Runtime client. The retrieval configuration used \texttt{overrideSearchType = HYBRID}, which allowed the knowledge base to combine lexical keyword matching and semantic vector retrieval. For each query, the system retrieved an initial set of 20 candidate evidence chunks. Each returned chunk included the original Bedrock retrieval rank, Bedrock retrieval score, page number, chunk identifier, source location, metadata, and retrieved text. Duplicate chunks were removed using the Bedrock chunk identifier when available; otherwise, the first portion of the text was used as a fallback deduplication key.

In the second phase, the deduplicated candidate chunks were passed to the Cohere reranking model. The reranker received the original query and the retrieved chunk texts as input and assigned a relevance score to each candidate chunk. The reranked output reordered the candidate chunks according to query-specific relevance. For the retrieval-result file, the top 20 reranked chunks were retained for each query to support later analysis of retrieval and reranking behavior. These outputs were saved in a structured CSV file containing the query identifier, query text, final reranked position, original Bedrock rank, Bedrock score, Cohere rerank score, page number, chunk identifier, source location, text preview, and full chunk text.

In the third phase, grounded answers were generated from the top-ranked retrieved evidence. For each query, the pipeline repeated the hybrid retrieval and reranking process and selected the top five reranked chunks as the final evidence context. These chunks were formatted as numbered source blocks containing page information, source location, and retrieved text. The answer-generation prompt instructed the model to use only the retrieved evidence, avoid outside knowledge, write in a formal academic style, and present one factual claim per sentence. The generation model was configured with a low temperature of 0.1 to encourage conservative and evidence-focused responses. If the retrieved evidence was insufficient, the model was instructed to explicitly state that the retrieved sources did not provide enough evidence.

In the final phase, the generated answers were evaluated using a claim-level grounding procedure. The evaluation module loaded the saved answer file and reranked chunk file without performing additional retrieval, reranking, or answer generation. Each generated answer was split into candidate factual claims using sentence-level segmentation, while short fragments, headings, table rows, and generic insufficient-evidence statements were removed. For each query, the saved reranked chunks were sorted by final rank and formatted into an evidence context. Each extracted claim was then evaluated against the retrieved evidence using Amazon Nova Pro as a judge model. The judge was instructed to return a structured JSON output containing a binary support decision, the best supporting source, and a brief explanation. A claim was marked as supported only if the retrieved evidence directly justified the full meaning of the claim. Claims with incomplete, vague, indirect, or partial support were treated as unsupported.

The pipeline generated three categories of output files. The first file stored hybrid retrieval and reranking results at the chunk level. The second file stored the generated answer for each query. The third set of files stored claim-level grounding judgments, query-level grounding accuracy, and overall grounding accuracy. This output design allowed the retrieval, reranking, generation, and evaluation stages to be inspected independently and supported systematic analysis of grounding reliability.

All intermediate outputs were stored as structured CSV files, including the retrieved and reranked chunks, generated answers, claim-level grounding judgments, query-level accuracy, and overall summary statistics. This design supports reproducibility by allowing each stage of the pipeline to be inspected independently.

%% file: Results.tex
\section{Results}

The proposed hybrid retrieval and reranking framework was evaluated using 25 natural-language queries related to biomedical natural language processing, transformer-based clinical language models, electronic health records, ICD code prediction, BioBERT, ClinicalBERT, and large language models in healthcare. For each query, the retrieval stage returned 20 candidate evidence chunks, resulting in 500 reranked evidence chunks across the complete evaluation set. During answer generation, the top five reranked chunks were used as the evidence context for each query. The generated answers were subsequently evaluated using claim-level grounding assessment. This result should be interpreted within the scope of the evaluated 25-query test set and the automated judge-based evaluation protocol.

Table~\ref{tab:overall_grounding_summary} summarizes the overall evaluation results. Across the 25 generated answers, the grounding evaluation process extracted 200 factual claims. The number of claims per query ranged from 1 to 12, with an average of 8.0 claims per query. All 200 claims were judged to be supported by the retrieved evidence, while no claims were marked as unsupported. Therefore, the overall claim-level grounding accuracy was 100.0\%. At the query level, all 25 queries achieved 100.0\% grounding accuracy, indicating that every factual claim extracted from each generated response was supported by the corresponding retrieved evidence.

\begin{table}[!t]
\centering
\caption{Overall claim-level grounding evaluation results.}
\label{tab:overall_grounding_summary}
\begin{tabular}{lc}
\hline
\textbf{Evaluation metric} & \textbf{Value} \\
\hline
Total evaluated queries & 25 \\
Total reranked evidence chunks & 500 \\
Reranked chunks per query & 20 \\
Chunks used for answer generation per query & 5 \\
Total extracted factual claims & 200 \\
Mean claims per query & 8.0 \\
Claim-count range per query & 1--12 \\
Supported claims & 200 \\
Unsupported claims & 0 \\
Overall grounding accuracy & 100.0\% \\
Queries with 100\% grounding accuracy & 25 of 25 \\
\hline
\end{tabular}
\end{table}

The evidence retrieved by the system was distributed across five source documents, as shown in Table~\ref{tab:source_distribution}. The largest proportion of reranked chunks came from the review paper on transformers and large language models in healthcare, which contributed 155 chunks, corresponding to 31.0\% of the retrieved evidence. The second-largest contribution came from the transformer models in biomedicine paper, which contributed 123 chunks, or 24.6\%. The lightweight transformers for clinical natural language processing paper contributed 87 chunks, or 17.4\%. The ICD prediction using ClinicalBERT paper contributed 70 chunks, or 14.0\%, while the BioBERT paper contributed 65 chunks, or 13.0\%. This distribution shows that the retrieval pipeline used evidence from multiple documents rather than relying on a single dominant source.

\begin{table}[!t]
\centering
\caption{Distribution of reranked evidence chunks across source documents.}
\label{tab:source_distribution}
\begin{tabular}{lcc}
\hline
\textbf{Source document} & \textbf{Chunks} & \textbf{Percentage} \\
\hline
Transformers and LLMs in healthcare review & 155 & 31.0\% \\
Transformer models in biomedicine & 123 & 24.6\% \\
Lightweight transformers for clinical NLP & 87 & 17.4\% \\
ICD prediction using ClinicalBERT & 70 & 14.0\% \\
BioBERT for biomedical text mining & 65 & 13.0\% \\
\hline
\textbf{Total} & \textbf{500} & \textbf{100.0\%} \\
\hline
\end{tabular}
\end{table}

The reranking stage substantially changed the initial retrieval order. Among the 500 retrieved chunks, only 44 chunks retained the same rank after reranking, corresponding to 8.8\% of the evidence set. For the top-ranked reranked chunks, the original Bedrock rank varied from 1 to 16, with a mean original rank of 6.0. Only 10 of the 25 final top-ranked chunks were originally ranked within the top three Bedrock retrieval results, while three final top-ranked chunks were originally ranked outside the top ten. These findings indicate that the reranking stage had a meaningful effect on evidence prioritization by reordering initially retrieved chunks before answer generation.

\begin{table}[!t]
\centering
\caption{Effect of reranking on retrieved evidence ordering.}
\label{tab:reranking_effect}
\begin{tabular}{lc}
\hline
\textbf{Reranking metric} & \textbf{Value} \\
\hline
Total reranked chunks & 500 \\
Chunks with unchanged rank & 44 \\
Percentage of unchanged ranks & 8.8\% \\
Mean original Bedrock rank of final top-ranked chunks & 6.0 \\
Minimum original Bedrock rank of final top-ranked chunks & 1 \\
Maximum original Bedrock rank of final top-ranked chunks & 16 \\
Final top-ranked chunks originally in Bedrock top 3 & 10 of 25 \\
Final top-ranked chunks originally outside Bedrock top 10 & 3 of 25 \\
\hline
\end{tabular}
\end{table}


The initial Bedrock retrieval score across all 500 retrieved chunks had a mean value of 0.501 with a standard deviation of 0.059. The Cohere rerank score across the same 500 reranked chunks had a mean value of 0.442 with a standard deviation of 0.242. Since these two scores are generated by different components, they should not be interpreted as directly comparable on the same absolute scale. However, within the Cohere reranking output, the 25 final top-ranked chunks, one for each query, received relatively high rerank scores, with a mean score of 0.801, a standard deviation of 0.098, and a range from 0.566 to 0.919. This suggests that the reranker assigned stronger relevance priority to the chunks selected at the top of the final evidence list.

\begin{table}[!t]
\centering
\caption{Retrieval and reranking score summary.}
\label{tab:score_summary}
\begin{tabular}{lcccc}
\hline
\textbf{Score type} & \textbf{Mean} & \textbf{Std.} & \textbf{Min.} & \textbf{Max.} \\
\hline
Bedrock retrieval score & 0.501 & 0.059 & 0.401 & 0.727 \\
Cohere rerank score & 0.442 & 0.242 & 0.016 & 0.919 \\
Top-ranked rerank score & 0.801 & .098 & 0.566 & 0.919 \\
\hline
\end{tabular}
\end{table}

The claim-level grounding results demonstrate that the generated answers were strongly aligned with the retrieved evidence. Because each generated answer was decomposed into individual factual claims, the evaluation provides a more fine-grained assessment than answer-level correctness alone. A generated response may contain both supported and unsupported statements; therefore, claim-level evaluation provides a stricter way to examine whether the response remains grounded in the retrieved evidence. In this experiment, all extracted claims were supported, indicating that the combination of hybrid retrieval, reranking, conservative prompting, and independent judge-model evaluation produced highly grounded responses for the selected biomedical NLP query set.

It is important to interpret these results as internal grounding accuracy rather than strict citation accuracy. The generated answers did not include explicit source labels in the final response. Therefore, the reported 100.0\% accuracy indicates that each factual claim was supported by at least one retrieved evidence chunk associated with the query. Strict citation accuracy would require each generated claim to include an explicit citation label and would further evaluate whether the cited source specifically supports that claim. Thus, the current experiment demonstrates strong evidence grounding, while future work should extend the framework to evaluate explicit sentence-level citation correctness.

Overall, the results show that the proposed hybrid retrieval and reranking framework produced evidence-grounded answers across all evaluated biomedical NLP and healthcare transformer queries. The reranking stage meaningfully changed the initial retrieval order, the retrieved evidence was distributed across multiple source documents, and the claim-level evaluation confirmed full support for all generated factual claims. These findings suggest that the framework can support reliable retrieval-augmented answer generation when the source document collection contains sufficient evidence for the target query. Future evaluation should include a larger and more diverse query set, human expert validation, explicit citation-label generation, and comparison against retrieval-only baselines to determine whether reranking improves both grounding accuracy and strict citation accuracy.

\subsection{Estimated Computational Time}

\begin{table}[!t]
\centering
\caption{Estimated execution time in Google Colab with NVIDIA A100 GPU.}
\label{tab:estimated_runtime}
\begin{tabular}{lc}
\hline
\textbf{Pipeline stage} & \textbf{Estimated time} \\
\hline
Environment setup and package loading & 1--3 min \\
Hybrid retrieval for 25 queries & 2--5 min \\
Cohere reranking of retrieved chunks & 3--6 min \\
Grounded answer generation & 5--8 min \\
Claim extraction and grounding evaluation & 8--12 min \\
CSV export and summary statistics & $<$1 min \\
\hline
\textbf{Estimated total runtime} & \textbf{20--35 min} \\
\hline
\end{tabular}
\end{table}

The experimental pipeline was executed in a Google Colab environment with an NVIDIA A100 GPU. For the evaluated setting of 25 queries, 500 reranked evidence chunks, and 200 claim-level grounding judgments, the estimated end-to-end wall-clock execution time was approximately 20--35 minutes. This estimate includes hybrid retrieval, reranking, answer generation, claim extraction, grounding evaluation, and result-file generation. It should be noted that the Amazon Bedrock and Cohere model calls were executed through external managed services; therefore, the total runtime was influenced primarily by API latency, network response time, model availability, and request processing limits rather than by GPU computation alone. The A100 GPU mainly supported the local notebook execution, data handling, and result processing steps. Although the experiment was conducted in a Google Colab A100 GPU environment, the main computational bottleneck was external model inference through Amazon Bedrock and Cohere APIs rather than local GPU computation.

%% file: Conclusion.tex
\section{Conclusion}

This study presented a hybrid retrieval and reranking framework for citation-aware retrieval-augmented generation. The proposed pipeline used Amazon Bedrock Knowledge Base for document ingestion, parsing, chunking, embedding generation, and evidence retrieval. Source PDF documents were stored in Amazon S3, embedded using Amazon Titan Text Embeddings V2, and indexed using Amazon OpenSearch Serverless. A hybrid retrieval strategy was then combined with reranking to prioritize the most relevant evidence chunks before answer generation.

The experimental results demonstrate that the proposed framework produced highly grounded responses for the evaluated biomedical natural language processing and healthcare transformer queries. Across 25 evaluation queries, the system retrieved and reranked 500 evidence chunks and generated responses using the top-ranked evidence. Claim-level grounding evaluation extracted 200 factual claims from the generated answers. All extracted claims were judged to be supported by the retrieved evidence, resulting in an overall grounding accuracy of 100.0\%. These findings indicate that the combination of hybrid retrieval, reranking, conservative prompting, and independent claim-level evaluation can improve the reliability of RAG-generated responses when sufficient evidence is available in the source document collection.

The reranking stage also played an important role in evidence prioritization. Only a small portion of the retrieved chunks retained the same rank after reranking, indicating that the reranker substantially changed the initial retrieval order. This suggests that reranking can help identify more directly relevant evidence from the candidate chunks returned by the initial retrieval stage. The distribution of retrieved evidence across multiple source documents further shows that the framework did not depend on a single document but was able to draw support from different papers within the knowledge base.

A key contribution of this work is the use of claim-level grounding evaluation rather than only answer-level assessment. By decomposing each generated response into factual claims, the evaluation process provided a more detailed measure of whether individual statements were supported by retrieved evidence. The use of a separate judge model for grounding assessment also helped reduce potential self-evaluation bias by separating the answer-generation model from the evaluation model.

However, the current study has several limitations. Because the present work was designed as a pilot-scale proof-of-concept study, the results should be interpreted as an initial feasibility evaluation rather than as a large-scale benchmark comparison. First, the evaluation was conducted on a relatively small set of 25 queries, which limits the generalizability of the findings. Second, the reported accuracy represents internal grounding accuracy rather than strict citation accuracy, because the generated answers did not include explicit source labels for each factual claim. Third, the grounding judgments were produced by an automated judge model and were not independently verified by human domain experts. Finally, the experiment did not include a direct comparison with retrieval-only, reranking-only, or non-hybrid baseline systems.

Future work should expand the evaluation to a larger and more diverse query set across multiple domains and document collections. Additional experiments should compare hybrid retrieval with purely lexical, purely vector-based, and retrieval-only baselines. The framework should also be extended to require explicit sentence-level source citations in the generated answers, allowing strict citation accuracy to be evaluated in addition to general grounding accuracy. Human expert review should be incorporated to validate automated grounding judgments and assess whether the generated responses are not only evidence-supported but also clinically and scientifically appropriate.

Overall, the results suggest that a hybrid retrieval and reranking approach can provide a strong foundation for reliable and evidence-grounded RAG systems. By combining structured document ingestion, semantic retrieval, reranking, controlled answer generation, and claim-level grounding evaluation, the proposed framework offers a reproducible approach for improving the trustworthiness of generated responses in biomedical and healthcare-related information retrieval tasks.

\section*{Reproducibility and Data Availability}

The materials required to reproduce the experimental workflow, including the query set, retrieved and reranked evidence outputs, generated answers, claim-level grounding evaluation results, query-level accuracy summaries, and overall summary files, are available in the project repository at: \url{https://drive.google.com/drive/folders/1QLNxniZ_Zxp42p8Y5fC_XQ6sWJzio9lP?usp=sharing}. The implementation was developed in Python using Google Colab and external managed services, including Amazon Bedrock Knowledge Bases, Amazon OpenSearch Serverless, Amazon Bedrock Runtime, and Cohere reranking. To support reproducibility, intermediate outputs were stored as structured CSV files so that retrieval, reranking, answer generation, and grounding evaluation could be inspected independently. API credentials, access keys, and service-specific authentication information are not included in the shared files for security reasons and should be supplied through secure environment variables when reproducing the experiments.

\section*{AI Use Disclosure}
The authors used large language models as experimental components within the proposed RAG pipeline. No AI tool was used to fabricate data or alter experimental results. The authors reviewed and verified the manuscript content, tables, and reported findings.